\documentclass[12pt, oneside]{article}   	
\usepackage{jheppub}
\usepackage{bbold}
\usepackage{amsmath}
\usepackage{amssymb}
\usepackage{amsfonts}
\usepackage{amsthm}
\usepackage{amsbsy}
\usepackage{array}
\usepackage{tikz}
\usetikzlibrary{arrows.meta}
\usetikzlibrary{snakes}
\tikzstyle{bag} = [align=center]
\usetikzlibrary{decorations.pathmorphing}

\def\is{\nspc  & \nspc  = \nspc  & \nspc }

\numberwithin{equation}{section}

\def\0{{(0)}}
\def\1{{(1)}}
\def\2{{(2)}}

\def\<{\langle }
\def\>{\rangle }
\def\[{\left[}
\def\]{\right]}

\newcommand{\bea}{\begin{eqnarray}}
\newcommand{\eea}{\end{eqnarray}}
\newcommand{\be}{\begin{equation}}
\newcommand{\ee}{\end{equation}}
\newcommand{\ba}{\begin{align}}
\newcommand{\ea}{\end{align}}

\newcommand{\tr}{\mbox{tr}}

\renewcommand{\epsilon}{\varepsilon}

   \makeatletter
  \let\over=\@@over \let\overwithdelims=\@@overwithdelims
  \let\atop=\@@atop \let\atopwithdelims=\@@atopwithdelims
  \let\above=\@@above \let\abovewithdelims=\@@abovewithdelims
\renewcommand\section{\@startsection {section}{1}{\z@}%
                                   {-3.5ex \@plus -1ex \@minus -.2ex}
                                   {2.3ex \@plus.2ex}%
                                   {\normalfont\large\bfseries}}

\renewcommand\subsection{\@startsection{subsection}{2}{\z@}%
                                     {-3.25ex\@plus -1ex \@minus -.2ex}%
                                     {1.5ex \@plus .2ex}%
                                     {\normalfont\bfseries}}

\newcommand{\Tr}{{\rm Tr}}

\def\Q{{\hat Q}}

\def\tr{{\rm tr}}
\def\be{\bea}
\def\ee{\eea}

\def\spc{\hspace{.5pt}}

\def\bea{\begin{eqnarray}}
\def\eea{\end{eqnarray}}
\def\is{\nspc  & \nspc  = \nspc  & \nspc }

\def\ba{\begin{eqnarray}}
\def\ea{\end{eqnarray}}

\def\nspc {} 

\def\tr{{\rm tr}}
\def\diag{{\rm diag}}
\let\pa=\partial
\def\sotimes{\raisebox{1pt}{\footnotesize $\otimes$}}

\def\J{{\cal J}}

\def\tQ{{\mbox{\tiny $Q$}}}
\def\tone{{\mbox{\tiny$1$}}}
\def\ttwo{{\mbox{\tiny$2$}}}
\enlargethispage{\baselineskip}

\newpage

\makeatletter
\def\@fpheader{\ }
\makeatother

\title{Towards A String Dual of SYK}
\author{\ \ Akash Goel and Herman Verlinde}
\affiliation{Department of Physics, Princeton University, Princeton, NJ 08544, USA}

\abstract{We propose a paradigm for realizing the SYK model within string theory. Using the large $N$ matrix description of $c<1$ string theory, we show that the effective theory on a large number $Q$ of FZZT D-branes in $(p,1)$ minimal string theory takes the form of the disorder averaged SYK model with $J \psi^{\spc p}$ interaction. The SYK fermions represent open strings between the FZZT branes and the ZZ branes that underly the matrix model. The continuum SYK dynamics arises upon taking the large $Q$ limit. We observe several qualitative and quantitative links between the SYK model and $(p,q)$ minimal string theory and propose that the two describe different phases of a single system.  We comment on the dual string interpretation of double scaled SYK and on the relevance of our results to the recent discussion of the role of ensemble averaging in holography.
}
\date{}

\begin{document}

\maketitle

\pagebreak

\addtolength\parskip{.75mm}

\addtolength{\abovedisplayskip}{1mm}
\addtolength{\belowdisplayskip}{1mm}
\section{Introduction}
\vspace{-1mm}

The SYK model is the prototype of a maximally chaotic quantum system with low energy dynamics given by near-AdS$_2$ gravity. It describes the quantum mechanics of $N$ Majorana variables $\psi_i$ governed by a non-linear interaction \cite{kitaevTalks,Sachdev:1992fk, Sachdev:2015efa,Maldacena:2016hyu}
\bea
\label{sykone}
S_{{\nspc }_{\spc \rm SYK}} \nspc \is\nspc  \int \nspc  d\tau\,  \Bigl( \sum_i  \psi_i \partial_\tau \psi_i -  H_{{\nspc }_{\rm \spc SYK} } \Bigr) \nonumber\\[-2.5mm]\\[-2.5mm]\nonumber
H_{{\nspc }_{\rm \spc SYK} } \is i^{p/2} \sum_{i_1\ldots i_p} J_{i_1\ldots i_p} \psi_{i_1}\psi_{i_2} \ldots \psi_{i_p}
\eea
with Gaussian random couplings $
\bigl\langle  J^2_{i_1\ldots i_p}\bigr\rangle \nspc  =\nspc  \J^2\, \frac{p!}{2p^2 N^{p-1}}$. The status of the SYK model as a candidate holographic dual of a near-AdS${}_2$ quantum gravity theory is supported by the maximal Lyapunov behavior of its OTOCs and by the presence of a dynamical Schwarzian goldstone mode \cite{kitaevTalks,Maldacena:2016hyu}. The dual theory has JT gravity as its low energy limit \cite{Almheiri:2014cka, Maldacena:2016upp, Engelsoy:2016xyb, Jensen:2016pah}, but little else is known about its dynamical content or its UV completion. The main obstacle to finding this UV complete bulk dual is that the SYK model itself has thus far not found its home within string theory. 

In this paper we present a proposed string realization of the SYK model. Our construction starts from minimal string theory \cite{Seiberg:2003nm, Seiberg:2004at, Maldacena:2004sn, Kutasov:2004fg} and follows the standard holographic paradigm based on open-closed string duality \cite{Maldacena:1997re, McGreevy:2003kb, Klebanov:2003km}. The main idea is to consider the worldvolume theory of a large number $Q$ of FZZT branes in $(p,1)$ string theory \cite{Fateev:2000ik, Teschner:2000md, Gaiotto:2003yb, Hashimoto:2005bf}. Using the large $N$ matrix description of minimal string theory, we show that the effective theory on the FZZT branes takes the form of the  SYK model with $J \psi^{\spc p}$ interaction. 
The SYK fermions represent open strings between the FZZT branes and the ZZ branes that underly the matrix description of the minimal string.  
The continuum SYK dynamics arises upon taking the large $Q$ limit. 

Perhaps surprisingly, we will find that, rather than the model \eqref{sykone} with given random couplings, our construction gives rise to the SYK model after disorder averaging
\bea
\label{syktwo}
S \nspc  \is  \int \nspc  d\tau\,  \sum_i  \psi_i \partial_\tau \psi_i 
-  \frac{N \J^2}{2p^2}\int \nspc  d^2\tau\,  G(\tau_\tone\nspc ,\tau_\ttwo)^p \\[1.5mm]
& & \quad G(\tau_1,\tau_2) \equiv 
\frac 1 N \sum_i \psi_i(\tau_1) \psi_i(\tau_2).
\eea
Hence the string theory seems to have no knowledge of the `microscopic' SYK Hamiltonian, but only of the ensemble average of SYK Hamiltonians.

\begin{figure}[t]
\begin{center}
\includegraphics[scale=.66]{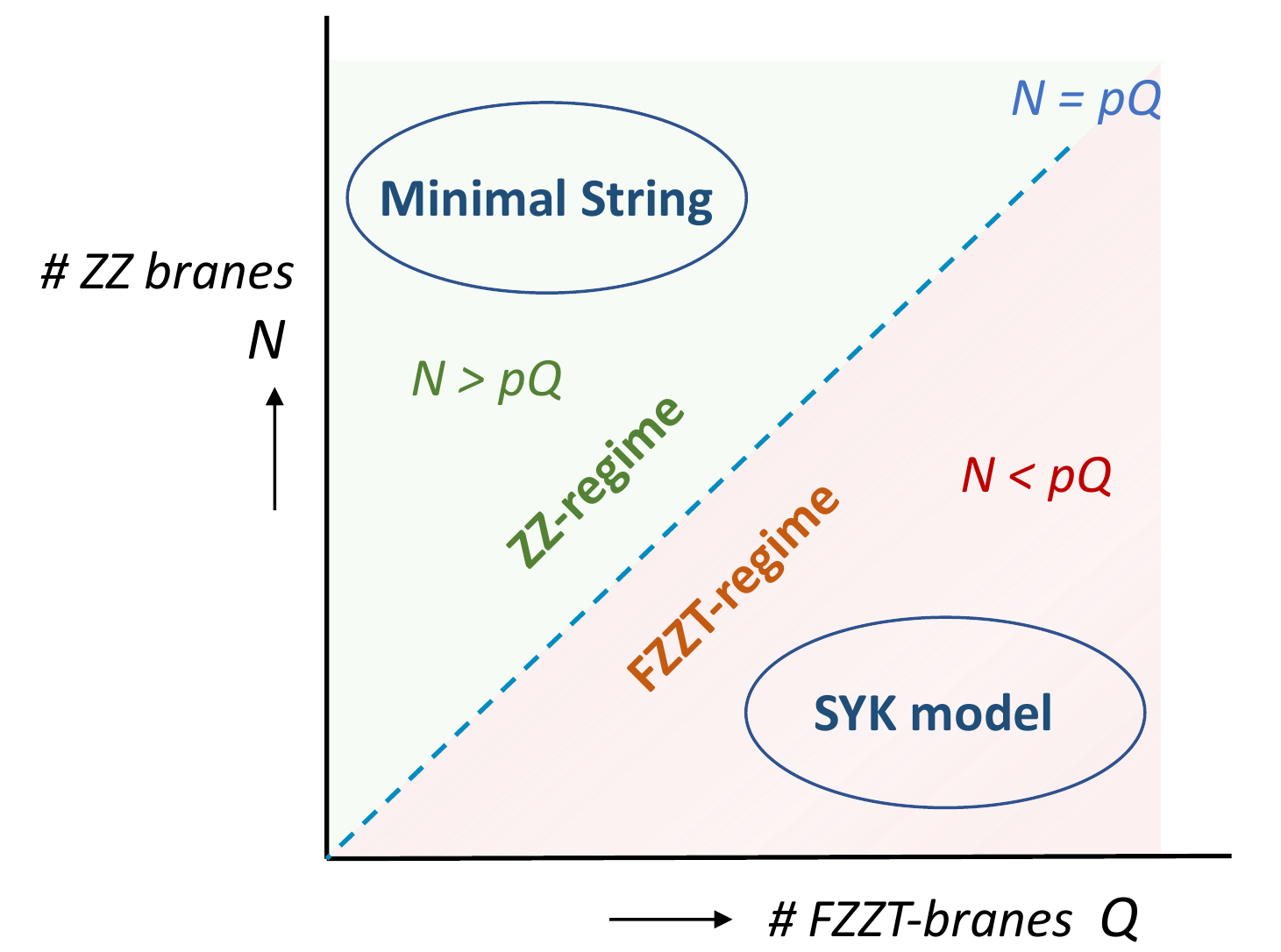}\qquad\qquad
\vspace{-3mm}
\end{center}
\caption{\label{fig:spectralcurve} The phase diagram of $(p,1)$ minimal string theory with $Q$ FZZT branes. The $(p,q)$ minimal string emerges in the regime where the number of ZZ-branes $N$ is much larger than $p \, Q$. The continuum SYK model with $\J \psi^p$ interaction resides in the regime when $p\, Q\gg N$. }
\end{figure}

Our derivation will reveal several qualitative and quantitative links between the SYK model with $\J \psi^{\spc p}$ interaction and $(p,q)$ minimal string theory and indicates that the two theories describe different phases of a single unifying theory labeled by three integers: the label $(p,1)$ of the minimal model, the number of ZZ-branes $N$, the number of FZZT branes $Q$.  The minimal string arises in the ZZ-regime where $N$ is taken much larger than $p\, Q$, whereas the SYK model emerges in the FZZT-regime with $p\, Q$ much larger than $N$.

The organization of this paper is as follows.  In section 2 we consider the worldvolume theory of $Q$ FZZT branes in $(p,1)$ minimal string theory. This system contains all $(p,q)$ minimal models. We then show that the partition function of the FZZT-ZZ open strings takes the form of a matrix version of the SYK model. In section 3 we describe how to take the continuum limit and obtain SYK quantum mechanics. In sections 4 and 5 we present the minimal string-SYK phase diagram and propose an interpretation of the double scaled SYK model \cite{Cotler:2016fpe, Berkooz:2018qkz, Berkooz:2018jqr, douglas-talk-kitp} as the worldsheet theory of a non-critical string. We end with some comments on how our results shed light on the role of ensemble averaging in holography.  In Appendix A and B contain a brief review of D-branes in minimal string theory, their representation in the dual two matrix model \cite{Douglas:1989dd,Daul:1993bg, Kazakov:2004du, Eynard:2002kg} and describe the geometry of the spectral curve of the $(p,q)$ minimal string.  Our presentation will be concise: a more detailed account of our results will appear elsewhere.

\pagebreak

\vspace{-1mm}

\section{SYK from the Two Matrix Model}
\vspace{-1.5mm}

In this section we consider $Q$ FZZT branes in $(p,1)$ minimal string theory. This system gives a general parametrization of the space of all $(p,q)$ minimal string theories. We use the matrix model description of minimal strings to show that the worldvolume of the FZZT branes contains fermionic degrees of freedom which are interpreted as FZZT-ZZ open strings. We then use a color-flavor transformation to show that their partition function takes the form of a matrix version of the SYK model.

\vspace{-1.5mm}

\subsection{FZZT-branes in the two matrix model}

\vspace{-1.5mm}
 The worldsheet of a minimal string is described by a 2D minimal model CFT coupled to Liouville theory. Appendix A contains a very brief decription  of the spectrum of operators and D-branes in this theory. The physical meaning and properties of the ZZ and FZZT branes in minimal string theory are most clearly exhibited via the dual matrix model perspective. 
Minimal string theory has a dual description as a double scaled two matrix model with partition function
\bea
Z(g,t) \nspc  \is\nspc  \int\!  dA\spc dB \, \exp\Bigl({-\Tr\bigl(V(A) +W(B) - AB\bigr)\Bigr)}  
\eea 
with $A$ and $B$ both $N\nspc \times\nspc  N$ matrices \cite{Daul:1993bg, Kazakov:2004du, Eynard:2002kg}. For the $(p,q)$ minimal string, the matrix potentials 
\bea
\label{expand}
V(A)\nspc  = \nspc \sum_n\spc g_n \spc A^n, \quad\ & &  \quad\ W(B)\nspc  =\nspc  \sum_k \spc t_k \spc B^k 
\eea
constitute polynomials of degree $p$ and $q$ and need to be tuned to a special $p/q$ multi-critical point.
In the old paradigm,  the Feynman diagrams of the double scaled matrix integral generate a dynamical triangulation of the minimal string worldsheet \cite{Ginsparg:1993is}.  In D-brane language, the matrix variables $A$ and $B$ represent open string between $N$ ZZ-branes \cite{McGreevy:2003kb,Klebanov:2003km, Seiberg:2003nm}. The double scaled matrix model aims to capture the full non-perturbative dynamics of the minimal string theory.

The duality between the two matrix model and the $(p,q)$ minimal string is most directly born out by considering the moduli space of the FZZT-brane as a function of the boundary cosmological constant $x= \mu_B$ (see Appendix A). This target space \eqref{fzztspec} coincides with the spectral curve of the matrix model \cite{Seiberg:2003nm}. The FZZT-brane at location $x$ is represented in the matrix model via the insertion of the determinant operator
\bea
\Psi(x)\nspc  \is \det(x -B)
\eea
This matrix description of the FZZT brane gives immediate insight into its worldsheet content. We can represent the determinant as a fermionic integral via
\bea
\label{FZZTferm} \det(x-B)= \int d\psi^\dagger d\psi\,
\exp\bigl({\psi^\dagger(x-B) \psi\bigr)}
\eea
The fermionic variables $\psi$ are naturally interpreted as representing the open strings that stretch between the FZZT brane and the $N$ ZZ branes that underly the matrix description of the minimal string \cite{Kutasov:2004fg}. The determinant formula can be generalized to $Q$ FZZT branes
\bea
\label{detferm}
\det\bigl(X_{\mbox{\nspc\tiny$Q$}}\spc \sotimes \, \mathbb{1}_{\mbox{\tiny$N$}}\nspc  -\nspc  \mathbb{1}_{\mbox{\nspc\tiny$Q$}}\spc \sotimes\, B\bigr)
\nspc  \is\nspc   \int \nspc   d\psi d\psi^\dag\spc \exp\Bigl({\psi^\dag\nspc (X_{\mbox{\nspc\tiny$Q$}}\spc \sotimes\,  \mathbb{1}_{\mbox{\tiny$N$}} -\mathbb{1}_{\mbox{\nspc\tiny$Q$}}\spc \sotimes\, B)\psi}\bigr)\quad\
\eea
with $X_{\mbox{\nspc\tiny$Q$}}$ an $Q\times Q$ matrix with eigenvalues $x_1,...,x_\tQ$. Hence the fermions have two indices
\bea
\psi_{ia}, \qquad i = 1,...,N, \qquad a=1,...,Q
\eea
indicating that they correspond to strings that stretch between $N$ ZZ-branes and $Q$ FZZT branes. We will call $i$ the color index and $a$ the flavor index.

Now suppose we turn off the potential $W(B)$ of the matrix model. This results in the matrix model of the $(p,1)$ minimal string. We can then choose to reintroduce $W(B)$ by placing by  $Q$ FZZT branes at suitable locations $x_s$, arranged such that we can equate
\bea
\label{vreplace}
\exp\bigl({-\Tr\, W(B)}\bigr)\nspc \nspc  \is
 \det\bigl(X_\tQ\, \sotimes\, \mathbb{1}_{\mbox{\tiny$N$}}\nspc  -\nspc  \mathbb{1}_\tQ\,\sotimes\, B\bigr).
\eea
This replacement amounts to using the Kontsevich parametrization of the matrix potential couplings $t_k$ in \eqref{expand}
in terms of the $Q\times Q$ matrix $X_\tQ =\diag(x_s)$ via~\cite{Kontsevich:1992ti}
\bea
t_k \is \frac{1}{k}\, \Tr\biggl( 
\frac{1}{X_\tQ^k}\biggr) 
\,= \, \frac{1}{k}\, \sum_{s=1}^Q \, \frac{1}{x_s^k}.
\eea
Note that for finite $Q\times Q$ matrix $X_\tQ$, the resulting potential $W(B)$ is not polynomial. However, we will be interested in taking the large $Q$ limit. In this case, the  Kontsevich parametrization is a one-to-one mapping from the space of potentials to the space of matrices $X_\tQ$.  The basic property that determines the dictionary is that both sides should lead to the same spectral curve.

\pagebreak

\subsection{Kontsevich matrix model from FZZT branes}

\vspace{-1.5mm}

We are led to study the matrix partition function
\bea
\label{zpq}
Z _{p\tQ}\nspc  \is\nspc  \int\!  dA\spc dB \, e^{\mbox{\footnotesize$-\Tr(V_p(A) - AB)$}}  \det\bigl(X_\tQ\spc \sotimes\, \mathbb{1}_{\mbox{\tiny$N$}} \nspc \nspc -\nspc \mathbb{1}_\tQ\, \sotimes\, B\bigr)\ \
\eea
In the double scaling limit,  this describes the partition function of $Q$ FZZT branes in $(p,1)$ minimal string theory. Taking the large $N$ and $Q$ limit and adjusting the locations $X_\tQ$, this system spans the complete space of all $(p,q)$ minimal models and their deformations.

It is instructive to briefly discuss the case $p=2$, following \cite{Maldacena:2004sn}.
The worldsheet of the $(2,1)$ minimal string is described by pure 2D topological gravity \cite{Witten:1989ig}. Its correlation functions compute the intersection numbers of the moduli space of Riemann surfaces \cite{Kontsevich:1992ti}. Inserting the $Q$ determinants amounts to turning on couplings $t_k$.
Since $V_2(A) =  \frac g 2 A^2$ is quadratic, we can integrate out $A$, producing a gaussian one-matrix model in $B$. Introducing fermionic variables via \eqref{detferm}, one can integrate out $B$ and obtain
\bea
Z_{2,\tQ} \nspc \is\nspc  \int\!  d\psi \spc d\psi^\dag\spc \exp\Bigl({\psi^\dag (X_{\nspc  \tQ}\, \sotimes \, \mathbb{1}_{\mbox{\tiny$N$}}) \psi - \frac g 2\Tr_{{\nspc }{N}}\nspc  ((\psi\spc \psi^\dag)^2)}\Bigr)
\eea
Using that  $\Tr_{{\nspc N}}\nspc ((\psi\spc \psi^\dag)^2) = \tr_Q\nspc ((\psi^\dag \psi)^2)$ with 
\bea
\label{psimatrices}
(\psi\spc \psi^\dag\nspc )_{ij}\nspc  = \nspc  \sum_{a=1}^Q \psi_{ia} \psi^\dag_{ja}\qquad 
(\psi^\dag\psi)_{ab}\nspc  = \nspc  \sum_{i=1}^N \psi^\dag_{ia} \psi_{ib}
\eea
and introducing a $Q\times Q$ Hubbard-Stratonovich matrix $\Sigma$, we can integrate out the fermions and rewrite the $Z_{2,\tQ}$ partition function as a $Q\times Q$ matrix integral
\bea
\label{prekonts}
Z_{2,\tQ} \nspc  \is\nspc   \int\! d\Sigma \; e^{\mbox{\footnotesize $-\frac 1 {2g} \tr( \Sigma^2)$}} \det\bigl(X_\tQ - \Sigma\bigr)^N
\eea
Taking a double scaling limit produces the famous Kontsevich matrix-Airy integral \cite{
Maldacena:2004sn, Kontsevich:1992ti}. 

In D-brane language the above manipulations have the following interpretation
\begin{enumerate}
\addtolength{\parskip}{-3mm}
\item write the $(p,1)$ partition function in terms of open strings $A, B$ between $N$ ZZ branes
\item introduce $Q$ FZZT branes at locations $X_\tQ\nspc  =\nspc  {\rm diag}(x_s)$
\item write their partition function in terms of fermionic ZZ-FZZT open string variables $\psi$
\item integrate out the ZZ open strings $A$ and $B$, producing a non-linear $\psi^4$ interaction 
\item integrate in FZZT open strings $\Sigma$ and integrate out the fermionic open strings
\addtolength{\parskip}{1.5mm}
\end{enumerate}
Performing these steps expresses the minimal string partition function in terms of the open strings on $Q$ FZZT branes. The $(2,q)$ minimal string partition functions are obtained by tuning the couplings $t_k$ to coincides with the higher critical points of the one matrix model.

\pagebreak

\subsection{SYK matrix model from FZZT branes}

\vspace{-1.5mm}

The above discussion directly generalizes to all integer values of $p$ \cite{Hashimoto:2005bf}. The $(p,1)$ minimal string worldsheet is described by topological gravity coupled to a topological minimal model. i.e. a twisted ${\cal N}\nspc =\nspc 2$ minimal CFT \cite{Dijkgraaf:1990dj}. This topological string theory has analogs of ZZ and  FZZT-branes \cite{Gaiotto:2003yb}. Upon taking the double scaling limit, equation \eqref{zpq} represents the partition function of $Q$ FZZT-branes in $(p,1)$ minimal string theory. The open string interpretation of all matrix variables described in the previous subsection carries over to this setting and one can perform the same steps. 

Starting with equation \eqref{zpq} for the partition function $Z_{p\Q}$, and after introducing the fermionic variables via \eqref{detferm}, one can first integrate out $B$ and then integrate out $A$ with the help of the resulting delta-function. This results in a fermionic integral 
\bea
Z _{p\Q}\nspc  \is\nspc  \int\!  \nspc  dA dB d\psi d\psi^\dag \,\exp\bigl(\psi^\dag(X_\tQ\,\sotimes\, \mathbb{1}_{\mbox{\tiny$N$}} \nspc \nspc  -\mathbb{1}_\tQ\, \sotimes\, B)\psi - \Tr_N(V_p(A) -AB)\Bigr)\nonumber \\[-2mm]\\[-2mm]\nonumber
\is\nspc  \int\nspc \nspc  d\psi \spc d\psi^\dag\spc \exp\Bigl({\psi^\dag (X_\tQ\, \sotimes \, \mathbb{1}_{\mbox{\tiny$N$}}) \psi - \Tr_{{\nspc }{N}}\nspc  V_p(\psi\spc \psi^\dag)}\Bigr)
\eea
 with a non-linear potential given by the trace of an $N\times N$ matrix $V_p(\psi\spc \psi^\dag)$. 
 
 As before, we can rewrite the potential as a trace of a $Q\times Q$ matrix by performing the {\it color-flavor transformation} \cite{Altland:2020ccq}
\bea
\Tr_{{\nspc N}}\bigl((\psi\spc \psi^\dag)^k\bigr) = \tr_{\nspc  Q}\bigl((\psi^\dag \psi)^k\bigr)
\eea
with $\psi\psi^\dag$ and $\psi^\dag\psi$ the matrices defined in \eqref{psimatrices}.

The partition function of the $Q$ FZZT strings then rearranges itself in the form of a matrix version of the SYK model
\bea
\label{matrixsyk}
Z _{p\tQ}\nspc \is\nspc  \int\nspc  d\psi\spc d\psi^\dag \exp\Bigl(\sum_{i=1}^N \psi_i^\dag X_\tQ \psi_i - {N \tilde\J^2} \,\tr_\tQ \spc G^{\spc p}\Bigr) \nonumber \\[-8mm]
\eea
with\\[-13mm]
\bea
G_{ab} \equiv \frac 1 N  \sum_{i=1}^N \psi^\dag_{ia} \psi_{ib} 
\eea
This expression should be compared with equation \eqref{syktwo} for the action of the SYK model after disorder averaging.
Here we only kept the highest order term of the potential $V_p(A)$ since all lower order terms are subleading in $1/N$ when expressed in terms of the new scaling variable $G_{ab}$. We already wrote its coupling constant in a suggestive way.

\clearpage

\section{Towards the Continuum SYK model}

\vspace{-1.5mm}

We have found that the partition function $Z_{p\tQ}$ of $Q$ FZZT branes in the matrix model formulation of $(p,1)$ minimal string theory looks like a discretized version of an SYK model. Next, we need to show that there exists a suitable choice of brane locations $X_\tQ$ such that we can take a continuum limit that reduces to SYK quantum mechanics. 

\vspace{-1.5mm}

\subsection{Non-commutative SYK}

\vspace{-1.5mm}

Comparing \eqref{matrixsyk} and \eqref{syktwo}, it is evident that the matrix $X_\tQ$ of FZZT brane locations should be thought of as a discretized time derivative. In this section, we will simply proceed and engineer the brane configuration accordingly. We will choose
\bea
\label{xqt}
X_\tQ\nspc \is\nspc   {\rm diag}\Bigl(\spc 2\spc \sin\frac{\raisebox{-.5pt}{\small $\pi s$}}{\raisebox{.5pt}{\small $Q$}}\Bigr)
\eea
with $s$ an integer running between $-Q/2$ and $Q/2$. (Here we assume $Q$ = even.) Taking the large $Q$ limit,  $\theta = \pi s/Q$ labels a continuous angle between $-\pi/2$ and $\pi/2$. In the next section we will argue that this special choice of brane locations has a natural geometric interpretation from the perspective of $(p,Q)$ minimal string theory. 

To make contact with SYK, we need to identify $Q\times Q$ matrices with bi-local functions of time. To this end we introduce $Q\times Q$ clock and shift matrices $U$ and $V$ with the property
\bea
U V = \zeta V U, \qquad  \zeta \spc = \spc e^{i \hbar} \qquad \hbar \equiv \frac{2\pi}{Q}.
\eea
$U$ and $V$ can be thought of as two non-commutative coordinates with the periodicity property $U^Q=V^Q=1$.
An arbitrary $Q\times Q$ matrix can be uniquely expanded as 
\bea
G \spc = \nspc  \sum_{m,n=1}^Q\nspc   G_{mn} U^m V^n
\eea
Writing
$U= e^{iu}$ and $V=e^{iv}$ with $u$ and $v$ periodic with $2\pi$, this decomposition provides a mapping from the space of $Q\times Q$ matrices to the space of functions on the non-commutative torus \cite{Connes:1997cr}. The matrix $G$ is mapped to the function $G(u,v)$, the trace becomes the integral and matrix multiplication is mapped to the star product  \cite{Connes:1997cr}
\bea
\tr_Q \to \hspace{-3mm} && \hspace{-3mm} \int\nspc  \frac{du dv}{2\pi \hbar} \qquad\ \ 
G\cdot G \ \to \  G* G \\[2mm]
G*G \is e^{\mbox{\footnotesize $i \hbar (\pa_u\pa_{\tilde{v}} -\pa_v\pa_{\tilde{u}})$}} G(u, v)\, G(\tilde{u},\tilde{v})_{|\mbox{\small ${{{}_{u=\tilde u}}\atop{{}^{v=\tilde v}}}$}}
\eea
Note that in the large $Q$ limit, this star product becomes an ordinary commutative multiplication of functions, provided that we restrict to functions that have a finite derivative as $\hbar \to 0$. We will assume that this physical restriction is justified.

Via the same dictionary, we can map $Q$ component vectors $\psi$ to functions of $u$ or $v$. We choose the latter option. Define the $Q$ basis vectors $|v\rangle$ via the relations
\bea
V|v\rangle \nspc \is\nspc  e^{iv} |v\rangle \qquad U|v\rangle = |v- \hbar\rangle
\eea
Decomposing an arbitrary vector $|\psi\rangle$ in this basis defines a function $\psi(v)$. So via the action on this basis, $U$ now acts like an off-diagonal matrix. It acts on the functions of $v$ via a finite shift $U \psi(v) = \psi(v + \hbar)$. Identifying $v$ with the time coordinate, we deduce that $U$ generates a small time shift. Defining $\langle \psi| = \frac 1 \hbar \int\! dv\, \langle v| \psi^*(v)$ we further note that
\bea
\label{inprod}
\langle \psi_1|\psi_2\rangle\nspc  \is \nspc  \frac 1 {\hbar}\int\nspc  dv\, \psi_1^*(v) \psi_2(v)
\eea

Writing $U$ in diagonal form $U =  \diag(e^{\frac{2\pi i s}{Q}})$ and comparing with \eqref{xqt}, we can  identify
\bea
i X_\tQ\nspc  \is \nspc  {U^{1/2}\nspc  - U^{-1/2}}
\eea
This $X_\tQ$-matrix acts on functions of $v$~via
\bea
\label{xqact}
i X_\tQ\spc \psi(v)\nspc  \nspc  \is\nspc   \Bigl(\textstyle \psi(v\nspc +\nspc  \frac{\hbar}{2}) \nspc  - \nspc  \psi(v\nspc -\nspc  \frac{\hbar }{2}) \Bigr)
\equiv   \hbar \hat\partial_v \psi(v) \ 
\eea
 confirming our anticipated conclusion that $X_\tQ$ acts as a discretized version of a time-derivative. 
 Combining this result with the formula \eqref{inprod} for the inner-product, we find that the kinetic term of the action indeed matches with the SYK kinetic term
\bea
\label{kinterm}
\psi^\dag \nspc i X_\tQ \psi \is \int \nspc \nspc  dv\, \psi^\dag \hat\partial_v \psi.
\eea

We are now ready to rewrite the matrix SYK action \eqref{matrixsyk} as a bi-local action in the emergent time variable, or equivalently, as a local action on the non-commutative torus. Introducing the two-point function $G$ and the fermion self-energy $\Sigma$ as two independent ($Q\times Q$ matrix) auxiliary degrees of freedom, we can write  \eqref{matrixsyk} in the following form
\bea
\label{ncsyk}
 S \nspc  \is \nspc   \int\nspc \nspc  \frac{d u dv}{2\pi}  \biggl(\nspc \sum_{i=1}^N \psi_i^\dag \bigl(\hat\partial_v\nspc  -\nspc  \Sigma\bigr) \psi_i\, +\, \nspc  N\Bigl( \Sigma * \spc G
+  \frac{\nspc \J^2\nspc }{2p^2\nspc } \spc G^{\spc *\spc p}\Bigr) \nspc \biggr)\ \ \ \
\eea
Here the kinetic term contains an implicit delta-function $2\pi \delta(u,v)$ that reduces the double integral to a single integral \eqref{kinterm} over $v$. 
The self-energy $\Sigma$ acts as a lagrange multiplier that imposes the identity
\bea
G(u,v) \nspc  \is\nspc   
\frac 1 N \sum_i \psi^\dag_i(u) \psi_i(v)
\eea
The only difference between \eqref{ncsyk} and the standard bi-local SYK lagrangian is that the derivative is defined via \eqref{xqact} and that the multiplication of functions proceeds via the star product.
In the large $Q$ limit, the star product becomes ordinary multiplication and $\hat{\partial_v}$ becomes an ordinary derivative. Both  these statements assume that all variables behave as sufficiently smooth functions of $u$ and $v$. In the continuum large $Q$ limit, the
 action \eqref{ncsyk} then looks identical to the standard SYK model \eqref{syktwo}, in units chosen such that $\beta = 2\pi$ and with Dirac fermions instead of Majorana fermions.
Hence the two-point function $G(u,v)$ is complex, but satisfies the non-local reality condition
\bea
G(u,v)^\dag \is G(v,u).
\eea
This suggests that we should fold up the torus by doubling the variables, so that the resulting continuum theory becomes local in $(u,v)$. We will take this step in section 5.

We deduce that, after taking the continuum limit $Q\to \infty$,  the fermionic variables satisfy the Dirac algebra $
\{\psi_i^\dag, \psi_j\} =  \delta_{ij}$
and generate a $2^N$ dimensional Hilbert space. Note that this Hilbert space dimension is exponentially large in $N$. Indeed, our original $N\times N$ matrices $A$ and $B$ do not have the interpretation of a random Hamiltonian as in \cite{Saad:2019lba, Blommaert:2019wfy}. Instead, they are open string wave functions.

\section{From Minimal Strings to SYK} \label{sec:mst-to-syk}

Our result that the partition function $Z_{p\spc Q}$ of $Q$ FZZT branes in the $(p,1)$ minimal string can be arranged to coincide with the SYK model establishes that the SYK model fits inside the larger class of theories scanned by moving the positions $x$ of the FZZT-branes. As emphasized above, this space of theories includes the $(p,q)$ minimal string theories. Hence it is natural to investigate how the two theories are connected. In this section we will indeed point to several qualitative and quantitative links between the SYK model and $(p,q)$ minimal string theory. 

\subsection{SYK and the $(p,q)$ spectral curve}

\vspace{-1.5mm}

Our first observation linking the SYK model and minimal string theory is that the special locations \eqref{xqt} that lead to the discrete time-derivative in the kinetic term in fact precisely coincide with a natural set of locations on the spectral curve of the $(p,q)$ minimal string.

The spectral curve of the minimal string describes the moduli space of the FZZT brane. It is parametrized by two variables $x$ and $y$, identified with the boundary cosmological constant $x=\mu_B$  and the derivative $y=\partial_{\mu_B} Z_D(\mu_B)$ of the disk partition function. The~two variables satisfy a relation of the form $F(x,y)=0$. For the $(p,q)$ minimal string, it reads
 \bea
 \label{spect}
T_p(y) -T_q(x) = 0
\eea 
with $T_p$ the Chebyshev polynomial of the first kind.  Equation \eqref{spect} specifies a $q$ sheeted branched cover of the $x$-plane and a $p$-fold cover of the $y$ plane. It can be uniformized by introducing the coordinate $z$ via
\bea
\label{solt} x=T_p(z),\qquad y=T_q(z).
\eea
This defines a genus $g_{pq}= {(p\nspc -\nspc 1)(q\nspc -\nspc 1)}/2$ complex curve  ${\cal M}_{pq}$ with $g_{pq}$ pinched A-cycles, as indicated in Fig 2.  The singular points of the curve ${\cal M}_{pq}$ coincide with the set of 
of solutions  $(x_{rs},y_{rs})$ to the ground ring relations $U_{p-1}(x)=U_{q-1}(y)=0$ given in \eqref{smallxrs}. 

\begin{figure}[t]
\begin{center}
\includegraphics[scale=.68]{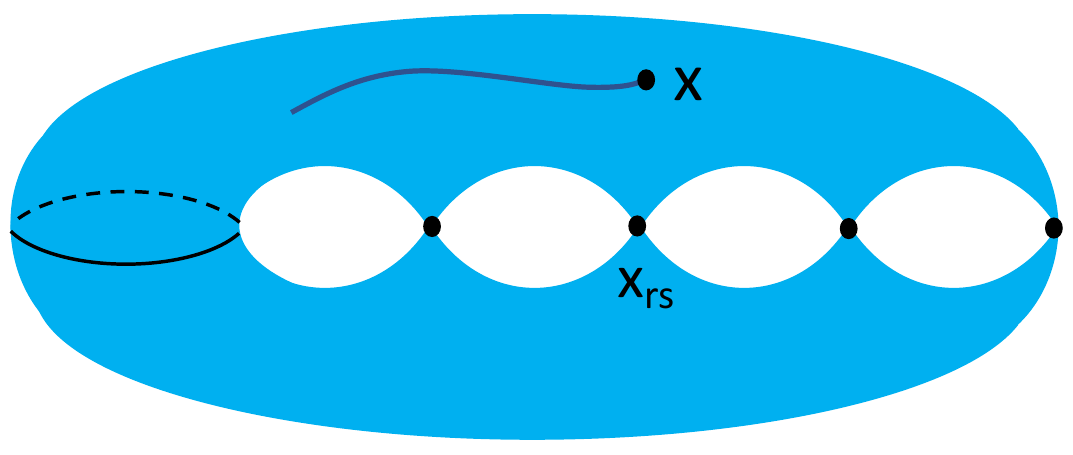}
\vspace{-2mm}
\end{center}
\caption{\label{fig:spectralcurve} Schematic depiction of a spectral curve ${\cal M}_{pq}$ (for $p=2$) with pinched cycles at $x_{rs}$ and FZZT brane located at $x$ }
\end{figure}

Comparing equations \eqref{xqt} with \eqref{smallxrs}, we observe that our special choice \eqref{xqt} for the matrix $X_\tQ$ has a very natural interpretation from the point of view of the minimal string: remarkably, it amounts to placing the FZZT branes precisely at the singular points \label{smallxrs} of a fiducial $(p,q)$ spectral curve ${\cal M}_{pq}$
\bea
\label{xqo}
X_\tQ\nspc \is\nspc  {\rm diag}(x_{rs}) = {\rm diag}\Bigl(\mbox{\small $(-)$}{}^r \spc 2\spc \cos\mbox{\large $ \frac{\pi p s}{q}$}\Bigr)
\eea
after performing a simple relabeling and provided we set $Q=q$. This correspondence points towards a deep geometric connection between the $(p,q)$ minimal string the discretized SYK model. However, as we will explain below, this connection can not be a direct physical correspondence. Instead, the two theories live in opposite corners of a phase diagram.

\subsection{FZZT brane correlation function}
\vspace{-2mm}

To better understand the relation between the SYK model and the minimal string, it will be useful to briefly consider correlation functions of FZZT-branes in minimal string theory with the help of the double scaled matrix model. The semi-classical expectation value of a single FZZT-brane operator $\Psi(x) = \det(x -B)$ assumes the form ~\cite{Maldacena:2004sn}  
\bea
 \bigl\langle  \Psi(x) \bigr\rangle\nspc  \is \frac{1}{\nspc \sqrt{\partial x(z)\nspc }}\; e^{\mbox{\small ${\frac 1 \hbar \int^x\nspc  y(x')dx'}$}} \equiv \psi_{cl}(x)
\eea
$\hbar$ incorporates the large $N$ scaling and $y(x)$ and $x(z)$ are specified by the spectral curve relations~\eqref{spect} and \eqref{solt}. We can think $\Psi(x)$ as the wave function of a particle with action 
\bea
S(x) \is -\int^x\nspc  \omega \qquad\qquad \omega \equiv y dx
\eea
$S(x)$ has non-trivial monodromy around the non-contractible cycles of the spectral curve. In particular, the periods of $\omega$ around the A-cycles measures how many branes are present at each singular point
\bea
\label{ZZcont}
\oint_{A_{r s}} \omega \is \hbar N_{rs}
\eea
Hence if $N_{rs}$ is non-zero, then $\omega$ and $\Psi(x)$ are both singular at  $x=x_{rs}$.

The semi-classical correlation function of $Q$ FZZT branes takes the form~\cite{Maldacena:2004sn} 
\bea
\label{fzztcorr}
\Bigl\langle \prod_{s=1}^Q \Psi(x_s) \Bigr\rangle \is \frac{\Delta(z)}{\Delta(x)} \prod_{s=1}^Q \psi_{\rm cl}(x_s)
\eea
The prefactor is the ratio of Vandermonde determinants of the relative locations of the branes parametrized by their $x$-location (denominator) or by their uniformizing coordinate $z(x)$ (numerator). The denominator vanishes whenever the $x$-locations of two FZZT branes coincide. The numerator, on the other hand, only vanishes if the two branes approach each other on the same sheet. In other words, the prefactor has a pole whenever two branes approach the same $x$-location, but are located on two different sheets. Since there are $p$ sheets, the correlator \eqref{fzztcorr} has $(p\nspc  -\nspc  1)(Q\nspc  -\nspc  1)$ poles as a function of $z_s$.

\subsection{Minimal String-SYK phase diagram}

\vspace{-1.5mm}

The spectral curve ${\cal M}_{p1}$ of $(p,1)$ minimal string theory is a $p$-fold cover of the $x$-plane.
We learned that the correlation function of $Q$ FZZT branes in this theory exhibits a pole whenever two branes approach the same $x$-location but on two different sheets of ${\cal M}_{p1}$. The residue of this pole has the physical interpretation of the number of ZZ-branes at this location. This indicates that each FZZT brane binds $(p-1)$ ZZ branes.

Our setup depends on three integers, the label $p$ of the minimal topological string, the number of ZZ-branes $N$ that make up the matrix model,  and the number of FZZT branes $Q$. The above argumentation suggests that we should distinguish three regimes
\bea
\label{npqrel}
a) \quad N\spc  &>& \, p\, Q 
\qquad\    \mbox{ZZ regime}\ \nonumber\\[.5mm]
b) \quad N\spc  \is \, p \, Q
\qquad\    
\mbox{critical}\   \\[.5mm]
c) \quad N\spc  &<&\, p \, Q 
\qquad\    \mbox{FZZT regime}\ \nonumber
\eea
The number $p\, Q$ represents the number of ZZ-branes bound to the $Q$ FZZT-branes. 
We propose that SYK and minimal string theory fit in on opposite regimes, as indicated in the phase diagram in Fig 1. 

In the super-critical regime a) we can take the large $N$ limit first. This case describes the physics of $Q$ FZZT branes in $(p,1)$ minimal string theory. Or if we take $Q$ large (but still sub-critical) and suitably adjust the locations of the branes, we can tune the effective matrix model potential so that we obtain the pure $(p,q)$ minimal string.  Hence all $(p,q)$ minimal string theories reside in the  ZZ-regime a). 

To obtain the continuum SYK model, on the other hand, we must take the large $Q$ limit first, keeping $N$ and $p$ initially finite.  Hence continuum SYK resides in the FZZT dominated regime c).  Another concrete way to see this is by considering the double scaled large $N$ limit with $N/p^2$ fixed. Since the interaction term $\J^2 G^{*\spc p}$ of the non-commutative SYK model contains a $p$-fold star-product, we should expect that deviations from the continuum result come in at order $\hbar p = p/Q$. So to get ordinary SYK, we need to send $p/Q\to 0$. This means that in the double scaling limit, we must also take $N/pQ = N/p^2 \times p/Q$ to zero.

The critical case b) separates the two regimes.  On the critical line, the $N$ ZZ-branes naturally bind to all $Q$ FZZT branes and fill out the lattice $(x_{rs}, y_{rs})$ of singular points of the $(p,Q)$ spectral curve ${\cal M}_{p\tQ}$. In the sub-critical case c), the ZZ-branes can no longer support the spectral curve and the semi-classical  large $N$ geometry breaks down.

 \clearpage

\section{SYK as a Non-Critical String} 
\vspace{-1.5mm}

Although they arise in different limits of our set up, there are several formal connections between the SYK model and minimal string theory. These become most apparent after taking the double scaling limit with $p^2/N$ fixed  \cite{Cotler:2016fpe, Berkooz:2018qkz, Berkooz:2018jqr, douglas-talk-kitp, Jia:2019orl}. Our results indicate that, similar to how the large $N$ limit of the matrix model captures tree level physics of the minimal string, double scaled SYK theory  describes the tree level physics of the string theory dual to SYK.

In the double scaled large $N$ limit with $p^2/N$ fixed,  the SYK dynamics is exactly captured by an effective Liouville theory \cite{Cotler:2016fpe}
\bea
\label{liouvilleone}
S[g]   \is \frac{\pi N}{p^2} \int\nspc   \frac{d u dv}{2\pi} \, \Bigl( \pa_u g\spc \pa_v g\, +\, 4\J^2 e^{2g(u,v)}\Bigr)
\eea
This effective lagrangian arises from \eqref{ncsyk} after first sending $q\nspc \to\nspc  \infty$, then integrating out the fermions and $\Sigma$ field, while writing $G(u,v)\nspc  = \nspc i\, {\rm sgn}(u,v)\spc \bigl(1\nspc  +\nspc   \frac 2 p \spc g(u,v)\bigr)$. Note that the lagrangian \eqref{liouvilleone} defines a 2D field theory in Lorentzian signature but appears in the SYK functional integral without a factor of $i$ in front. 
In our case, the Liouville field $g(u,v)$ is complex, but satisfies the reality condition $
g(u,v)^\dag \nspc =\nspc  g(v,u)$.
Hence instead of defining the model as living on the torus, we can orientifold the torus into a mobius strip and introduce two separate Liouville fields via
\bea
g_+(u,v) \nspc \is\nspc  g(u,v), \qquad \quad
g_-(u,v) = g(v,u).
\eea
This doubles up the lagrangian into a sum 
\bea
\label{sumaction}
 S[g] \nspc \is \nspc  S[g_+] + S[g_-]
\eea
of two Liouville lagrangians  with complex central charge\footnote{This result follows from the more standard formula
$c_\pm = 1\nspc  + 6Q_\pm^2$ with $Q_\pm\nspc =\nspc  b_\pm\nspc  +\nspc  1/{b_\pm}$ by taking
$b_\pm\nspc  = e^{\pm {i\pi}/{4}}\beta$. The double scaled SYK model exhibits $U_q(SU(1,1))$ quantum group symmetry with $q\nspc  =\nspc  e^{-\pi \beta^2}$ real, whereas standard Liouville theory has $U_q(SL(2,\mathbb{R}))$ symmetry with $q=e^{i\pi b^2}$ a complex phase.}
\bea
\label{complexc}
c_\pm\nspc  \is\nspc  13 \nspc  \pm\nspc  i\spc 6(\beta^2\nspc  -\nspc  \beta^{-2}), \qquad \beta^2\nspc = \nspc  \frac{\spc p^2}{2\pi N}
 \eea
The two complex central charges add up to $c_+ + c_- = 26.$ This result looks coincidental at first. Taken more seriously, however, it provides a direct hint that the bi-local effective theory of the double scaled SYK model should be viewed as a 2D worldsheet string theory.

Accordingly, the sum \eqref{sumaction}  of two bi-local actions \eqref{liouvilleone}
should be treated gravitational theory subject to Virasoro constraints. Indeed, although \eqref{liouvilleone} looks like the lagrangian of a 2D local quantum field theory, all evidence indicates that the 2D dual to the SYK model does not have a local stress tensor.  This formal similarity between double scaled SYK  and non-critical string theory motivated us to look for a possible realization of SYK via D-branes in (generalized) minimal string theory and matrix models.

Another hint of a connection between SYK and minimal string theory is found by considering the spectral density. For the double scaled SYK model with Majorana fermions, it can be written in the following form \cite{Cotler:2016fpe, Berkooz:2018jqr, Berkooz:2018qkz}
\bea
\rho(x) \is \frac{{\cal N}}{\sqrt{1-x^2}} \prod_{r,s} \bigl(x-X_{rs}\bigr) 
\eea
with  ${\cal N}$ some overall normalization, $x$ the SYK energy and
\bea
\label{largexrs}
X_{rs} = (-)^r \cosh(\pi \beta^2 s)
\eea
with $r=0,1$ and $s\in \mathbb{Z}_+$. The formula \eqref{largexrs} should be compared with equation \eqref{smallxrs} for the critical points of the spectral curve ${\cal M}_{pq}$. Again we see that double scaled SYK has features of a generalized minimal string theory with imaginary coupling $b^2 = i\beta^2$.

Liouville CFTs with complex central charge of the form \eqref{complexc} look somewhat unfamiliar\footnote{They naturally appear as boundary theories of $SL(2,\mathbb{C})$ CS theory and pure $dS_3$ gravity \cite{Cotler:2019nbi}. This hints at a possible 3D gravity perspective on the double scaled SYK model. }. Their appearance also seems unexpected given that we obtained the SYK model from a large $q$ limit of a minimal string theory with a $c<1$ worldsheet CFT. However, one can write the lagrangian in a somewhat more familiar form of a sine-Liouville theory, as a deformation of JT gravity, or as a $c\nspc  =\nspc  1$ string theory in a time dependent background.

As a concrete special case,
the action of the self-dual theory with $\beta^2\nspc  =\nspc  p^2/2\pi N\nspc  =\nspc  1$ and real central charge $c_+ = c_- = 13$, can be rewritten in the form of a non-critical $c=1$ string worldsheet CFT with a sine-Liouville interaction
\bea
\label{ktaction}
S = \vspace{-3mm} & \frac{\raisebox{2pt}{$1$}}{\raisebox{-4pt}{$2\pi$}}&\nspc \nspc  \int \nspc  \Bigl((\partial X)^2\nspc  + (\partial\phi)^2\nspc \nspc  +\nspc  \mbox{\large $\frac{Q}{4}$} {\cal R} \phi\nspc  +\nspc  \mu e^{\alpha\phi} \sin({X}/R)\Bigr)
\eea
with $Q = 2$, $\alpha = 2$ and $R=1/2$.
With these parameters the action \eqref{ktaction} splits as sum of two $c=13$ Liouville actions $S[g_+] + S[g_-]$ with $g_\pm = \phi \pm i X_\pm$.
From the target space perspective, the sine-Liouville term describes a time-dependent tachyon condensate. If we assume $X$ is periodic,  \eqref{ktaction} is T-dual to the worldsheet theory of compactified $c\!  =\!  1$ string theory at the Kosterlitz-Thouless point with a condensate of vortices.  The appearance of vortices is linked to the presence of non-singlet states in $c\!=\!1$ matrix quantum mechanics \cite{Gross:1990ub}, or in D-brane language, the presence of FZZT branes. To extract the target space physics, one needs to include the backreaction of the FZZT branes. It would be interesting to connect these ideas with earlier proposed dual descriptions of strings in a 2D black hole space-time \cite{Kazakov:2000pm}.  More generally, it is worth exploring how the SYK model can be used to elucidate the target space physics of non-critical string theory, and vice versa.

\vspace{-.5mm}

\section{Conclusion}
\vspace{-1.75mm}

We have presented a framework for realizing the SYK model within string theory. Our proposal fits within the standard paradigm of holographic string duality. We start with a closed string theory and identify its spectrum of D-branes; in our case these are the ZZ and FZZT branes of minimal string theory. We then take the number of branes to infinity and consider the dynamics of the corresponding open strings. The open strings divide up into three sectors, those between the ZZ-branes, between the FZZT branes and those between the ZZ and FZZT branes. The latter type are the pre-cursor to the SYK fermions. 
 
The properties of the dual string theory depend on which type of brane dominates. Minimal string theory resides in the regime in which the ZZ branes proliferate. The open-closed string dictionary follows the usual 't Hooft paradigm: the ZZ open strings diagrams span the closed string worldsheet. The continuum SYK model arises in the opposite regime in which the number $Q$  of FZZT branes is taken to infinity. The branes create a time crystal that in the large $Q$ limit becomes a continuum emergent time direction. The dynamics of the fermionic ZZ-FZZT open strings organizes itself in the form of an SYK action. 

We have argued that the worldsheet theory of the string dual of the SYK model looks similar to that of minimal string theory, as summarized in the below table
\bea
\begin{array}{cc} \mbox{MST}\qquad\qquad &\qquad \qquad\mbox{SYK}\qquad\\[1.5mm]
{c_\pm\! =\nspc 13\nspc \pm\nspc 6(b^2 \! +\nspc 1/b^{2})} \qquad\qquad &\qquad\qquad  {c_\pm\! =\nspc 13\nspc \pm\nspc i 6(\beta^2 \! -\nspc 1/\beta^{2})}\qquad\\[1.5mm]
{b^2 = p/q, \ \  \ \ N\to \infty} \qquad\qquad &\qquad\qquad {\beta^2 = p^2/2\pi N, \ \ \ q\to \infty}\qquad
\end{array}\quad
\eea
The ratios $p/q$ and $p^2/N$ play the role of a worldsheet coupling constant on their respective side, while $1/N$ has its usual role as governing the strength of the string interactions.

\smallskip

We end with mentioning three conceptual open questions.
 
A somewhat surprising aspect of our proposed string realization of the SYK model is that the string theory does not describe the model with given couplings $J_{i_1i_2...i_p}$ but rather produces the SYK model after disorder averaging. Of course, this conclusion is inevitable given that our proposed minimal string dual only depends on three numbers, $N$, $p$ and $q$ -- so the string theory clearly can have no knowledge of the specific microscopic realization of the SYK Hamiltonian. Still it is usually string theory that gives the most detailed window into the UV physics. In fact, one could argue that this is still the case in our situation. In particular, if we take our non-commutative intermediate description seriously, it provides a particular UV regulator of the short distance physics of the SYK model. It also hints at new types of dualities, such as the color-flavor transformation and  $p$\,--\spc$q$ duality of  minimal string theory, that would look quite mysterious from the pure SYK point of view.

The above comment about disorder averaging appears directly relevant to the ongoing discussions about whether low dimensional holography relates gravitational bulk theories to specific quantum systems or ensembles of quantum systems. Starting with \cite{Saad:2019lba, Marolf:2020xie}, specific implementations of the recent ideas connect the appearance of wormhole geometries in the bulk gravity theory to the $1/N$ expansion of a random matrix theory, in which the random matrix $M$  is identified with the Hamiltonian of the boundary theory and $N$ coincides with the dimension of its Hilbert space. In our matrix model, on the other hand, $N$ is just given by the number of ZZ branes on the minimal string side or the number of fermions on the SYK side. So our matrix is much smaller and our $1/N$ corrections are therefore much larger than the ones considered in \cite{Saad:2019lba, Marolf:2020xie}. Our $1/N$ corrections correspond to contributions from higher worldsheet topologies rather than to target space wormholes.

Perhaps the most radical feature of our proposal, is that time is emergent. Continuous time only arises after taking the strict large $q$ limit. Some earlier hint that time in SYK should be viewed as possibly discrete are found in the chord diagram expansion of the double scaled SYK theory \cite{Berkooz:2018jqr, Berkooz:2018qkz}, where time evolution is most effectively captured by means of a transfer matrix and the energy spectrum runs over a finite range, suggesting an interpretation as a quasi-energy of a system with a discrete time evolution.

Our story is far from finished. Given the central place of the SYK model as a prototype of low dimensional holography with the same dynamical properties as a quantum black hole, it is clearly important to find its connection with string theory. Our proposal aims to take a concrete step in that direction.

\section*{Acknowledgements}
\vspace{-1mm}

We thank Juan Maldacena, Thomas Mertens, Vladimir Narovlansky, Joaquin Turiaci, Robbert Dijkgraaf and Mengyang Zhang for valuable discussions and comments. The research of HV is supported by NSF grant number PHY-1914860.

\appendix 
\section{D-branes in Minimal String Theory}
\vspace{-1.5mm}

We briefly describe D-branes in minimal string theory, following \cite{Seiberg:2003nm,Seiberg:2004at}. The worldsheet of a minimal string is described by a 2D minimal model coupled to Liouville theory. 
Minimal CFTs \cite{Belavin:1984vu} are labeled by coprime integers~$(p,q)$, have central charge $c = \nspc  1\nspc -\nspc  6(b-1/b)^2 < 1$ with $b^2 \nspc  =\nspc  p/q$, and a finite number of
primary operators ${\cal O}_{r,s} \nspc =\nspc  {\cal O}_{p-r,q-s}$. 
The integers $r$ and $s$ run from 1 to $p\nspc  -\nspc  1$ and $q\nspc -\nspc  1$, so in total there are  ${(p-1)(q-1)}/2$ primary operators. The minimal model is coupled to a Liouville CFT with action
\bea
\label{Laction} S \is\nspc  {1\over4\pi}\int\nspc  d^2z\spc
\bigl((\partial\phi)^2- 4\pi \mu e^{2 b \phi} \bigr)
\eea
and central charge
$ c\nspc =\nspc 1\nspc + 6Q^2\nspc  > 25$ with $Q\nspc =\nspc  b +  1/b$. Physical operators take the form $\hat{{\cal O}}_{rs}={\cal L}_{rs} {\cal O}_{rs}e^{2\,
 \alpha_{rs}\, \phi }$
with ${\cal L}_{rs}$ a combination of Virasoro generators  and $2\nspc  \alpha_{rs}\nspc =\nspc Q \nspc - \nspc  rb\nspc - \nspc s/b$. 

The physical operators 
form an associative, commutative ring isomorphic to the fusion ring of the minimal model CFTs.  Denoting the ground ring generators by
\bea
X =   \hat{{\cal O}}_{12}\quad & & \quad Y = \spc \hat{{\cal O}}_{21}
\eea
the ring relations take the form
\bea
U_{p-1}(Y) \nspc  \is\nspc  U_{q-1}(X) = 0, \qquad \quad
T_p(Y)- T_q(X)
= 0,
\eea
with $T_p$ and $U_p$ Chebyshev polynomials of the first and second kind.

There are two kinds of D-branes in minimal theory. The FZZT branes form a continuous family parametrized by the boundary cosmological constant $\mu_B$ 
\bea
\label{deltaS}
S_B \is \mu_B \oint e^{b\spc \phi}
\eea
It describes a Cardy state $|\sigma\rangle$ associated with a Virasoro representation with conformal dimension $\Delta={1\over 4} \sigma^2 +\frac 1 4 {Q^2}$.
Here $\mu_B$ and $\sigma$ are related via
\bea
\mu_B = \cosh \pi b \sigma  \equiv x
\eea
The variable $x$ can be thought of as the position of the endpoint of the FZZT brane.
In $\phi$ space, FZZT branes have semi-infinite extent and stretch from $\phi\nspc =\nspc -\infty$ to $b\phi \simeq -\log\mu_B = -\log x$.
Following \cite{Seiberg:2003nm,Seiberg:2004at,Maldacena:2004sn}, we also define a $y$ variable via the disk amplitude $Z_{{\nspc }{D}}$ 
\bea
y(x) \is \partial_x Z_{{\nspc }{D}} \quad \leftrightarrow \quad Z_{{\nspc }{D}} = \int^x\nspc  y(x') dx'\ 
\eea
For the $(p,q)$ minimal string, $Z_D = \cosh\pi\sigma/b$. 

The $x$ and $y$ variables satisfy the relation 
 \bea
 \label{fzztspec}
F(x,y)= T_p(y) -T_q(x) = 0.
\eea 
This equation specifies a $q$ sheeted branched cover of the $x$-plane and a $p$-fold cover of the $y$ plane. It can be uniformized by introducing the coordinate $z$ via
\bea
\label{solz} x=T_p(z),\qquad y=T_q(z).
\eea
Equation \eqref{fzztspec} defines a genus ${(p\nspc -\nspc 1)(q\nspc -\nspc 1)}/2$ complex curve  ${\cal M}_{pq}$ with ${(p\nspc -\nspc 1)(q\nspc -\nspc 1)}/2$ pinched A-cycles, as indicated in Fig 1.  This curve can be viewed as the target space geometry probed by the FZZT brane. 

ZZ branes are localized in the strong coupling region. Their boundary state can be written as the difference of two FZZT boundary states~\cite{Martinec:2003ka}
\bea
\label{ZZbs}
|r,s\rangle = |\,\sigma_{rs} \rangle - |\, \sigma_{r,-s}\rangle
\eea
where $\sigma_{rs} = i({r/b} + sb)$ labels the Cardy state associated with the $\Delta_{rs}$ degenerate Virasoro representation of the minimal worldsheet CFT. The ZZ branes define eigenstates of the ground ring generators
\bea
\label{greigen}
 X|r,s\rangle\nspc  \nspc \is\nspc   x_{rs}
 |r,s\rangle, \quad\ \ \quad 
 Y|r,s\rangle = \spc  y_{rs}
 |r,s\rangle.\\[-9mm]\nonumber
\eea
with\\[-9mm]
\bea 
\label{smallxrs}
x_{rs} \nspc \is\nspc  (-1)^r \cos \mbox{\large$\frac{\pi p s}{q}$ }, \ \qquad \mbox{\small $s=1,...,q\nspc -\nspc 1$}\nonumber\\[-2mm]\\[-2mm]\nonumber
y_{rs} \nspc \is\nspc  (-1)^s \cos \mbox{\large$\frac{\pi q r}{p}$ }, \ \qquad \mbox{\small $r=1,...,p\nspc -\nspc 1$}
\eea
the set of solutions to the ground ring relations $U_{p-1}(x)=U_{q-1}(y)=0$. These locations $(x_{rs},y_{rs})$ correspond to the singular points of the curve ${\cal M}_{pq}$. Hence the ZZ branes are situated at the pinched cycles.

\bibliographystyle{ssg}
\bibliography{Bibliodraft}

\newpage
\section*{Addendum to ``Towards a String Dual of SYK'': Mapping between the Matrix Model SYK and the Non-Commutative SYK model}

\vspace{-1.5mm}

In this Addendum we give a more detailed description of the mapping between the Matrix Model SYK presented in section 2.3 equation \eqref{matrixsyk} and the non-commutative SYK model presented in section 3.1.\footnote{We thank Sumit Das, Animik Ghosh and Antal Jevicki for pointing out that the dictionary between the two modifications of the SYK model is less straightforward than implicated in the main text. The purpose of this addendum is to show that basic statements in sections 3 and 4 remain valid, but that the dictionary requires the introduction of an additional gaussian random matrix, as explained below.} To set notation and keep the discussion in this addendum self-contained, we will first present a non-commutative version of the SYK model, which will both allow for a matrix description and that manifestly reduces to the continuum SYK model in the large $Q$ limit. This part of the discussion in the addendum overlaps with section 3.2 in the main text. We will then present a concrete dictionary between this Non-Commutative SYK theory and the Matrix SYK model.

\subsection{Matrices and the non-commutative torus}

\vspace{-1.5mm}

To make contact with SYK, we need to identify $Q\times Q$ matrices with bi-local functions of time. To this end we introduce $Q\times Q$ clock and shift matrices $U$ and $V$ with the property
\bea
U V = \zeta V U, \qquad  \zeta \spc = \spc e^{i \hbar} \qquad \hbar \equiv \frac{2\pi}{Q}.
\eea
$U$ and $V$ can be thought of as two non-commutative coordinates with the periodicity property $U^Q=V^Q=1$.
An arbitrary $Q\times Q$ matrix can be uniquely expanded as 
\bea
\hat{G} \spc = \nspc  \sum_{m,n=1}^Q\nspc   G_{mn} U^m V^n
\eea
Writing
$U= e^{iu}$ and $V=e^{iv}$ with $u$ and $v$ periodic with $2\pi$, this decomposition provides a mapping from the space of $Q\times Q$ matrices to the space of functions on the non-commutative torus \cite{Connes:1997cr}. In the non-commutative geometry notation, the matrix $\hat{G}$ is represented by a function $\hat{G}(u,v)$. The trace becomes an integral and matrix multiplication is mapped to the star product  \cite{Connes:1997cr}
\bea
\tr_Q \to \hspace{-3mm} && \hspace{-3mm} \int\nspc  \frac{du dv}{2\pi \hbar} \qquad\ \ 
G\cdot G \ \to \  G* G \\[2mm]
G*G \is e^{\mbox{\footnotesize $i \hbar (\pa_u\pa_{\tilde{v}} -\pa_v\pa_{\tilde{u}})$}} G(u, v)\, G(\tilde{u},\tilde{v})_{|\mbox{\small ${{{}_{u=\tilde u}}\atop{{}^{v=\tilde v}}}$}}
\eea
Note that in the large $Q$ limit, this star product becomes an ordinary commutative multiplication of functions, provided that we restrict to functions that have a finite derivative as $\hbar \to 0$. We will assume that this physical restriction is justified.

Via the same dictionary, we can map $Q$ component vectors $\psi$ to functions of $u$ or $v$. We choose the latter option. Define the $Q$ basis vectors $|v\rangle$ via the relations
\bea
V|v\rangle \nspc \is\nspc  e^{iv} |v\rangle \qquad U|v\rangle = |v- \hbar\rangle
\eea
Decomposing an arbitrary vector $|\psi\rangle$ in this basis defines a function $\psi(v)$. So via the action on this basis, $U$ now acts like an off-diagonal matrix. It acts on the functions of $v$ via a finite shift $U \psi(v) = \psi(v + \hbar)$. Identifying $v$ with the time coordinate, we deduce that $U$ generates a small time shift. Defining $\langle \psi| = \frac 1 \hbar \int\! dv\, \langle v| \psi^*(v)$ we further note that
\bea
\label{inprod}
\langle \psi_1|\psi_2\rangle\nspc  \is \nspc  \frac 1 {\hbar}\int\nspc  dv\, \psi_1^*(v) \psi_2(v)
\eea

\medskip

\subsection{Non-commutative SYK}

\vspace{-1mm}

We now introduce the non-commutative SYK model. This model has a matrix description and in the large $Q$ limit reduces to the continuum SYK model. In the next section, we describe the mapping between the NCSYK model and the Matrix SYK model.

To write the kinetic term, we introduce the following discrete time derivative 
\bea
\label{xq}
i X_\tQ\nspc  \is \nspc  {U^{1/2}\nspc  - U^{-1/2}}
\eea
This $X_\tQ$-matrix acts on functions of $v$~via
\bea
\label{xqact}
i X_\tQ\spc \psi(v)\nspc  \nspc  \is\nspc   \Bigl(\textstyle \psi(v\nspc +\nspc  \frac{\hbar}{2}) \nspc  - \nspc  \psi(v\nspc -\nspc  \frac{\hbar }{2}) \Bigr)
\equiv   \hbar \hat\partial_v \psi(v). \ 
\eea
Plugging this into the formula \eqref{inprod} for the inner-product, we obtain a good candidate for the kinetic term of the discretized SYK model
\bea
\label{kinterm}
\sum_{i=1}^N \langle \psi | \nspc i X_\tQ |\psi\rangle \is \sum_{i=1}^N \int \nspc \nspc  dv\, \psi_i^\dag(v) \hat\partial_v \psi_i(v).
\eea

To write the interaction term of the NCSYK model, we use the fact that to a set of $N$ fermionic functions with finite fourier expansions 
\bea
\psi_i(v) = \sum_{n=1}^Q \psi_{in} e^{inv} \qquad \quad \psi_i^{\dagger}(u) =\sum_{m=1}^Q \psi_{im} e^{imu} 
\eea
we can associate a $Q\times Q$ matrix $ \widehat{G}$ via\footnote{
Note that this matrix $\widehat{G}$ is {\it not} the same as the matrix defined by the dyadic sum $\widetilde{G} = \frac{1}{N}\sum_i |\psi_i\rangle\langle \psi_i |$. We will return to this point in the next section.}
\bea
\label{hatG}
\widehat{G} \is \frac{1}{N} \sum\limits_i 
\psi_i^{\dagger}(U) \psi_i(V) 
\eea
The total Non-Commutative SYK action is now defined as follows
\bea
\label{ncsykm}
S_{\rm NCSYK} \! \is\! \sum_{i=1}^N \langle \psi_i | X_Q | \psi_i \rangle + {\cal J}^2 \tr(\spc {\widehat{G}^p} )
\eea
with  $\widehat{G}$ defined in \eqref{hatG}.

The NCSYK action can be rewritten as a bi-local action in the emergent time variable, or equivalently, as a local action on the non-commutative torus. Introducing the two-point function $G$ and the fermion self-energy $\Sigma$ as two independent ($Q\times Q$ matrix) auxiliary degrees of freedom, we can write  \eqref{ncsykm} in the following form
\bea
\label{ncsyk}
 S_{\rm NCSYK} \nspc  \is \nspc   \int\nspc \nspc  \frac{d u dv}{2\pi}  \biggl(\nspc \sum_{i=1}^N \psi_i^\dag \bigl(\hat\partial_v\nspc  -\nspc  \Sigma\bigr) \psi_i\, +\, \nspc  N\Bigl( \Sigma * \spc G
+  \frac{\nspc \J^2\nspc }{2p^2\nspc } \spc G^{\spc *\spc p}\Bigr) \nspc \biggr)\ \ \ \
\eea
Here the kinetic term contains an implicit delta-function $2\pi \delta(u,v)$ that reduces the double integral to a single integral \eqref{kinterm} over $v$. 
The self-energy $\Sigma$ acts as a lagrange multiplier that imposes the identity
\bea
\label{guv}
G(u,v) \nspc  \is\nspc   
\frac 1 N \sum_i \psi^\dag_i(u) \psi_i(v)
\eea

The non-commutative SYK model allows for a relatively straightforward continuum limit. The only difference between \eqref{ncsyk} and the standard bi-local SYK lagrangian is that the derivative is defined via \eqref{xqact} and that the multiplication of functions proceeds via the star product.
In the large $Q$ limit, the star product becomes ordinary multiplication and $\hat{\partial_v}$ becomes an ordinary derivative. Both  these statements assume that all variables behave as sufficiently smooth functions of $u$ and $v$. In the continuum large $Q$ limit, the
 action \eqref{ncsyk} then looks identical to the standard SYK model \eqref{syktwo}, in units chosen such that $\beta = 2\pi$ and with Dirac fermions instead of Majorana fermions.
Hence the two-point function $G(u,v)$ is complex, but satisfies the non-local reality condition
\bea
G(u,v)^\dag \is G(v,u).
\eea

\vspace{-1.5mm}

\subsection{Matrix SYK}

\vspace{-1.5mm}

Now let us return to our matrix SYK model \eqref{matrixsyk} derived from considering $Q$ FZZT branes in minimal string theory. Comparing the matrix action \eqref{matrixsyk}  with the non-commutative action \eqref{ncsykm}, we see that the matrix $X_\tQ$ of FZZT brane locations will have to be arranged such that it takes the form of a discretized time derivative \eqref{xq}. Here we will simply proceed and assume that we can engineer the brane configuration accordingly.
Writing $U$ in diagonal form $U =  \diag(e^{\frac{2\pi i s}{Q}})$ and comparing with \eqref{xqt}, we see that we should place the brane locations at
\bea
\label{xqt}
X_\tQ\nspc \is\nspc   {\rm diag}\Bigl(\spc 2\spc \sin\frac{\raisebox{-.5pt}{\small $\pi s$}}{\raisebox{.5pt}{\small $Q$}}\Bigr) 
\eea
with $s$ an integer running between $-Q/2$ and $Q/2$. (Here we assume $Q$ = even.) Taking the large $Q$ limit,  $\theta = \pi s/Q$ labels a continuous angle between $-\pi/2$ and $\pi/2$. In section \ref{sec:mst-to-syk} we will argue that this special choice of brane locations has a natural geometric interpretation from the perspective of $(p,Q)$ minimal string theory.

We use the definitions $
\tilde{\psi}_i(u) = \langle u|\tilde{\psi}_i\rangle$ and $\tilde{\psi}^\dag_i(v) = \langle\tilde{\psi}_i|v\rangle,$
and write $U$ in diagonal form $U =  \diag(e^{\frac{2\pi i s}{Q}})$. The tildes distinguish the fermionic variables from the corresponding variables in NCSYK. Comparing with \eqref{xqt}, we can  identify
\bea
iX_Q= \int \! du dv \, \delta(u\!-\!v) |v\rangle \langle u| (U^{1/2}\! - U^{- 1/2})
\eea
Using this notation, we can now rewrite the Matrix Model SYK action \eqref{matrixsyk} as
\bea
S_{\rm MMSYK} \is \sum_i  \langle \tilde{\psi}_i | X_Q | \tilde{\psi}_i \rangle \spc + N \spc {\cal J}^2 \tr( \spc {\widetilde{G}^p} ), \qquad \qquad 
\eea
where $\widetilde{G}$ denotes the dyadic sum
\bea
\label{tildeg}
\widetilde{G} \is \frac{1}{N} \sum_{i=1}^N \,
| \tilde{\psi}_i \rangle \langle \tilde{\psi}_i | 
\eea
The above matrix SYK action looks very similar to the non-commutative SYK action given in equation \eqref{ncsykm}. Note, however, that the two bilocal fields $\widehat{G}$ defined in \eqref{hatG} and $\widetilde{G}$ defined in \eqref{tildeg} are not automatically equal. This is most directly seen by computing the matrix elements between $U$ and $V$ eigenstates
\bea
\langle v| \widetilde{G} |u\rangle = \frac 1 N\, \tilde\psi^\dag(u) \tilde\psi(v), \quad & & \quad
\langle v| \widehat{G} |u\rangle = \frac 1 N\, \psi^\dag(u) \psi(v) \langle u| v\rangle,
\eea
with $G(u,v)$ given in \eqref{guv}. In the next subsection, we will show that the two theories are nonetheless equivalent, provided one uses an appropriate mapping between observables. 


\vspace{-1.5mm}

\subsection{Mapping between Matrix SYK and Non-Commutative SYK}

\vspace{-1.5mm}

In the previous subsection, we denoted the Matrix Model SYK variables $\tilde{\psi}_i$ with extra tildes to indicate the fact the MMSYK variables are different from the NCSYK variable $\psi_i$. As we will now show, the two are related via a linear field redefinition
\be \label{psi-redefinition}
|\tilde{\psi}_i\rangle\, = \sum_k Z_{ik}(U) |{\psi}_k\rangle \qquad \qquad 
\langle \tilde{\psi}_i|\;  =  \sum_\ell\,  \langle{\psi}_\ell|\, Z^\dag_{\ell i}(V) 
\ee
where $Z_{ik}(U)$ is a suitably chosen random unitary transformation that acts both on the color and flavor indices
\be
\sum_k Z_{ik}(U) Z^\dagger_{kj}(U) = \delta_{ij} \otimes \mathbb{1}_{Q}\\[-4mm]\nonumber
\ee
We will see that this field redefinition is necessary and sufficient to make contact between the matrix model SYK and non-commutative SYK theories.

First we note that the redefinition \eqref{psi-redefinition} leaves the kinetic term invariant
\be\sum_i {\langle \tilde{\psi}_i | X_Q | \tilde{\psi}_i \rangle  = \sum_i \langle \psi_i | X_Q | \psi_i \rangle}
\eea
since $X_Q$ only depends on $U$ and thus commutes with the unitary tranformations $Z_{ik}(U)$. 
The bilocal operator in the interaction term, however, transforms non-trivially under \eqref{psi-redefinition}
\bea 
\widetilde{G} = \frac 1  N \sum\limits_i | \tilde{\psi}_i \rangle \langle \tilde{\psi}_i |
\is  \frac 1 N
\sum\limits_{i,k,\ell}
Z_{ik}(U)| \psi_k \rangle \langle \psi_\ell | Z^{\dagger}_{\ell i}(V)\\[-10mm]\nonumber
\eea
\vspace{-2mm}
so that
\bea
\langle u | \widetilde{G} | v \rangle \is  \frac 1 N \sum\limits_{k,\ell} 
\psi^\dagger_k(u) \psi_\ell(v) \sum_i Z^\dagger_{\ell i}(v) Z_{ik}(u) \\[-4mm]\nonumber
\eea
We will now pick the matrix $Z_{ik}(U)$ to be a {\it random matrix} selected from a Gaussian ensemble such that
\be
\sum_i \, \langle  Z^\dagger_{\ell i}(v) Z_{ik}(u) \rangle \is {\delta_{k\ell}}
{\langle u| v\rangle}\, = \, \delta_{k\ell} \,  e^{\frac i \hbar u v}
\ee
Here and below, $\langle ... \rangle$ denotes the disorder average over the gaussian unitary matrix ensemble.

With this choice of ensemble, we can now equate
\bea
\langle u | \widetilde{G} | v \rangle \is \frac{1}{N} \sum_i \tilde\psi_i^\dag(u) \tilde \psi_i(v) \\
\is \frac 1 N \sum\limits_{k,\ell} 
\psi^\dagger_k(u) \psi_\ell(v) \sum_i \langle Z^\dagger_{\ell i}(v) Z_{ik}(u)\rangle \nonumber \\
\is \frac 1 N \sum\limits_{k} 
\psi^\dagger_k(u) \psi_k(v)   \,  {\langle u| v\rangle}\; = \; \langle u | \widehat{G} | v\rangle
\eea
We abbreviate this relation as
\bea
\langle \spc \widetilde{G} \spc \rangle \is \widehat{G}
\eea
More generally, by large $N$ factorization of the gaussian matrix model, we have
\bea
\langle\,  \widetilde{G} \,\widetilde{G} \, \ldots \, \widetilde{G}\, \rangle \is
\langle\spc  \widetilde{G}\spc \rangle \,
\langle\spc  \widetilde{G}\spc \rangle \, \ldots \,  \langle\spc \widetilde{G}\spc \rangle  \, = \,\widehat{G}\,\widehat{G} \, \ldots \, \widehat{G}
\eea

Hence, with this additional element of the dictionary, we can identify correlation functions in the non-cummutative SYK theory with correlation functions of the matrix SYK theory, via the dictionary
\bea
\bigl\langle \, \widetilde{G} \ldots \widetilde{G}\, \bigl\rangle_{\rm MMSYK} \is \Bigl\langle \int [d{\tilde\psi}]\,  e^{-S_{\rm MMSYK}[\tilde\psi]}\, \widetilde{G} \ldots \widetilde{G} \, \Bigr\rangle \\[2mm]
\is \int[ d\psi] \, e^{-  \langle {\psi}_i| X_Q|{\psi}_i\rangle } \langle e^{- N {\cal J}^2 \tr(\widetilde{G}^p)}\, \widetilde{G}\ldots \widetilde{G} \rangle \\[2mm]
\is
\int[ d\psi] \, e^{-  \langle {\psi}_i| X_Q|{\psi}_i\rangle }  e^{- N {\cal J}^2 \tr(\widehat{G}^p)} \, \widehat{G}\ldots \widehat{G} \\[2mm]
\is \int [d{\psi}]\,  e^{-S_{\rm MMSYK}[\psi]}\, \widehat{G} \ldots \widehat{G} \, = \, 
\bigl\langle \, \widehat{G} \ldots \widehat{G}\, \bigl\rangle_{\rm NCSYK}
\eea
Here in the first line, the brackets on the right-hand side denote the disorder average over the $Z$ gaussian matrix model. In the second line, we use the fact that the integration measure and kinetic term are both invariant under the unitary redefinition \eqref{psi-redefinition} of the $\psi$ variables. In the next line, we again make use of large $N$ factorization to equate
\bea
\bigl\langle e^{-{\cal J}^2  \tr( \spc {\widetilde{G}^p} )} \bigr\rangle =  e^{-{\cal J}^2  \tr \spc {\langle \widetilde{G} \rangle^p} }  = e^{-{\cal J}^2 \tr(\spc \widehat{G}^p )} .
\eea

\end{document}